# Second harmonic generation with plasmonic metasurfaces: direct comparison of electric and magnetic resonances

Rohith Chandrasekar[1], Naresh K. Emani[1], Alexei Lagutchev[1], Vladimir M. Shalaev[1], Cristian Ciracì[2,3], David R. Smith[3], and Alexander V. Kildishev[1,*]

[1]*School of Electrical and Computer Engineering and Birck Nanotechnology Center, Purdue University, West Lafayette, IN 47907, USA*
[2]*Center for Metamaterials and Integrated Plasmonics, Department of Electrical and Computer Engineering, Duke University, Box 90291, Durham, NC 27708, USA*
[3]*Istituto Italiano di Tecnologia (IIT), Center for Biomolecular Nanotechnologies, Via Barsanti, I-73010 Arnesano, Italy*
[*]*kildishev@purdue.edu*

**Abstract:** Plasmonic resonances in metallic nanostructures have been shown to drastically enhance local electromagnetic fields, and thereby increase the efficiency of nonlinear optical phenomena, such as second harmonic generation (SHG). While it has been experimentally observed that enhanced fields can significantly boost SHG, to date it proved difficult to probe electrical and magnetic resonances in one and the same nanostructure. This however is necessary to directly compare relative contributions of electrical and magnetic components of SHG enhancement. In this paper we report an experimental study of a metasurface capable of providing electrical and magnetic resonant SHG enhancement for TM polarization. Our metasurface could be engineered such that the peak frequencies of electrical and magnetic resonances could be adjusted independently. We used this feature to distinguish their relative contributions. Experimentally it was observed that the magnetic resonance provides only 50% as much enhancement to SHG as compared to the electric resonance. In addition aligning both resonances in frequency results in conversion efficiency of 1.32 x $10^{-10}$.

## 1. Introduction

The advent of nanofabrication opened entirely new ways for designing optical materials, by virtue of the freedom to generate meta-atoms smaller than the wavelength with tailored electromagnetic responses, e.g., plasmonic resonances. With greatly improved spatial control of bulk parameters, such as the permittivity and permeability, we are now able to design optical metamaterials with properties unavailable in nature [1]. Applications of metamaterials include negative refractive index materials [2], invisibility cloaking [3], and phase-gradient metasurfaces [4] to name a few. Fundamentally, plasmonic resonances in optical metamaterials allow for light manipulation, concentration and enhancement of electromagnetic fields in the near-field [5]. This property has had an especially significant impact in the area of nonlinear optics.

Nonlinear optical phenomena are extremely useful in many applications, such as wavelength conversion, generation of ultra-short pulses, optical signal processing and ultrafast switching, but are inherently weak processes [6]. One such nonlinear phenomenon is second harmonic generation (SHG) – a second-order nonlinear process in which two photons at λ are annihilated and a photon at λ/2 is created, all in one single act [7]. SHG can be observed in non-centrosymmetric materials with non-zero second-order nonlinear susceptibility $\chi^{(2)}$. Though metals are centrosymmetric and do not possess a $\chi^{(2)}$ nonlinearity, symmetry is broken at interfaces and SHG is allowed. As a result, the nonlinear response from the metal surface is highly dependent on the surface morphology, and can be tailored by structuring the metal film [8]. Furthermore, due to its nonlinear dependence on the electromagnetic field, SHG efficiency can be substantially increased in environments that provide field enhancement. Optical metamaterials offer a platform for achieving this two-stage enhancement of second order nonlinear processes, through increased symmetry-breaking via nanopatterning of metal films and by concentrating electromagnetic fields with optically resonant nanostructures [9,10]. Efficient wavelength converters at nanometer length scales are required for various integrated optics applications as well as nonlinear optical microscopy. Importantly, such nanometer-thick devices are not subject to strict phase-matching conditions, allowing for low-dispersion and broadband wavelength conversion.

The enhancement provided by electric and magnetic resonances to SHG have been studied in metallic structures, such as nanoantennas [11-18], core-shell nanoparticles [19-21], split-ring resonators [22-28], as well as other unique systems [29-33].One of the first experimental studies of SHG in nanostructures with magnetic resonances was by Klein et al., where they studied SHG from split ring resonators (SRR) [22]. Since SRRs exhibit a magnetic and electric resonance for two different polarizations, Klein compared the effects of the resonances on SH efficiency. Fabricating two samples, one with the magnetic resonance at 1500 nm and the other one with its electric counterpart at the same wavelength, Klein observed that the magnetic resonance had a much stronger effect on SHG in SRRs. In general SHG signal generated in nanostructures is weak and often strongly varies from point to point at the surface due to sample nonuniformity inevitable in real life fabrication processes. To further complicate matters, if electric and magnetic resonances are excited by different polarizations, the respective surface near-field amplitudes will be quite different as well. As a result direct comparison of SHG efficiency among different samples or even different areas of the same is often difficult. In fact our experimental results further confirm just that.

Such difficulty also becomes apparent from the subsequent experiment performed by the same group where they measured SHG from the complementary SRR structure [25]. According to the generalized Babinet's principle, the magnetic and electric fields interchange with respect to the SRR - hence the magnetic dipole moment for the standard SRR will become an electric dipole moment for the complementary structure. Maximum SH output for this structure was observed at the electric resonance, contrary to their previous experiment. Since then, other works have been published on enhanced SHG from nanostructures, and specifically using structures with enhanced quality factors in order to achieve higher SH conversion efficiencies [12,18]. Some experimental works have used resonances only at the harmonics [33], while others have also tailored and mode matched multiple resonances in nanostructures to achieve higher conversion efficiencies [11,17,32].

However all comparative studies thus far have looked at electric and magnetic resonances in separate structures, which could introduce different levels of roughness and structural defects and therefore lead to disparate results [34]. Hence, as far as SHG nanostructures are concerned it would be of interest to experimentally study the system possessing both electric and magnetic resonances in one structure for the same polarization of fundamental light. In this way by fabricating such structure with electrical and magnetic resonances detuned by frequency it becomes possible to directly measure the ratio of their respective SHG contributions. This is accomplished by recording the emission spectrum from one spot on the sample eliminating the need to switch samples and polarizations of fundamental light.

In this work, for the first time to our knowledge, we demonstrate SHG by a metasurface arranged as a single-period array of coupled composite metal-dielectric nanostrips. This metasurface has been shown to support both symmetric and antisymmetric displacement current flow, hence exhibiting electric and magnetic resonances respectively for transverse magnetic (TM)- polarized light [35]. These structures have been termed as *metamagnetics* due to their extraordinary optical magnetic response, $\mathrm{Re}(\mu)<0$. We have already studied these structures simulating their nonlinear optical properties and effects of roughness. The latter have been explored experimentally as well [36,37]. Therefore, we deemed this metasurface to be well-suited for studying the interplay of electric and magnetic resonances. By specifically tailoring the electric and magnetic resonances in our design we can tune the electric resonance on or off the SH of the magnetic resonance. By doing so we can probe both resonances and get a deeper understanding of the enhancement they provide to the nonlinear process.

## 2. Design Optimization

The 3D schematic, cross section and SEM image of the two metasurfaces used in this study are shown in Figures 1a-1c. The unit cell of the metasurface was made of composite alumina-silver strips with trapezoidal cross-section. Each strip was composed of 35 nm thick silver strips with a variable alumina spacer layer between them to tune the magnetic resonance. A 5 nm base and top layer of alumina were deposited as well. The structure had a base width of 205 nm and top width of 125 nm, with a period of 310 nm, as shown in Figure 1b. Metasurfaces were fabricated on glass substrates with 15 nm Indium Tin Oxide (ITO)-coated substrates, using standard electron beam lithography techniques (Vistec VB6 UHR). Silver and alumina layers were deposited using an electron beam evaporator. After liftoff procedures, a 6 nm layer of silicon dioxide was deposited to protect silver from oxidation.

The experimental and simulated transmission spectra for TM polarized normal incident illumination are shown in Figure 1d. Linear spectra were acquired using a spectroscopic ellipsometer (V-VASE, J.A. Woolam Co.), and simulated spectra were acquired using a commercial finite-element frequency domain software (COMSOL Multiphysics). The permittivities of alumina and silica were taken to be 2.59 and 2.12 respectively, and a Drude-Lorentz fit to Johnson-Christy experimental data was used for the permittivity of silver [38]. A 15 nm thick ITO layer was used, with permittivity retrieved from ellipsometric data. A slight red shift in experimental curves compared to simulated can be due to roughness in the structure, as we have shown previously [37].

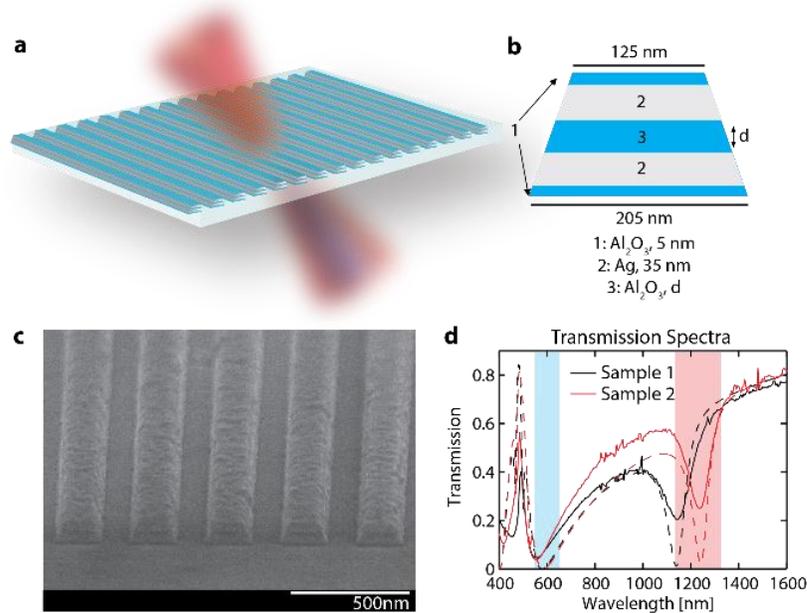

Fig 1. (a) 3D schematic of the metasurface consisting of coupled silver nanostrips gratings, (b) schematic of grating cross section, d = 19.5 nm for Sample 1 and 15 nm for Sample 2, (c) SEM image of metasurface, (d) Experimental (solid) and simulated (dashed) transmission spectra for Sample 1 and 2.

Both metamagnetic samples were designed to have an electric resonance at 570 nm, so that the matching fundamental wavelength was 1140 nm; the first sample (Sample 1) had a magnetic resonance at 1140 nm, and the second (Sample 2) had the resonance at 1240 nm. While the magnetic resonance of Sample 1 was matched with the fundamental of the electric resonance, the magnetic resonance of Sample 2 was slightly detuned in order to probe the contribution from each resonance individually. We achieved broad tunability in the magnetic resonance wavelength by simply changing the thickness of the alumina spacer layer between the silver nanostrips (~20 nm resonance shift/nm change in spacer). Alumina spacer layer thicknesses were 19.5 nm and 15 nm for Samples 1 and 2 respectively. Sample 2 provided the fairest comparison of enhanced SHG due magnetic and electric resonances, since they were exposed to the same roughness and structural defects. Sample 1 was used as a control to ensure that no other spectral features were observed when resonances were aligned. It also allowed us to observe any further enhancement that could have been achieved due to mode matching of the resonances. The red and blue regions in Figure 1d depict the excitation and emission wavelength ranges, respectively. While these measurements have been conducted at normal incidence, the resonances are still exhibited under 45° illumination, which is the angle at which second harmonic signal is collected. Please see the Appendix for field maps of the resonances and transmission spectra under 45° illumination.

## 3. Experiments

While this metasurface can generate second harmonic signals in both reflection and transmission, our interest is in the transmitted signal, which will be presented here. The SHG characterization setup consisted of a Ti:Sapphire (SpectraPhysics Spitfire) femtosecond laser fed into an Optical Parametric Amplifier (TOPAS, Light Conversion) to generate 100 fs pulses at 1kHz at the fundamental wavelengths, in this case λ=1140nm-1300nm. Please see the Appendix for detailed measurements of pump pulse widths. The TM-polarized IR-wavelength signal was isolated using a Glan-Thompson polarizer and a 1064nm long pass filter (Semrock BLP01-1064R-25), focused by a 15mm bi-convex lens onto the sample at 45° incidence. A filter stack consisting of colored glass filters (Thorlabs, KG-3) and extended hot mirrors (Edmund Optics 46386) was used to isolate the transmitted second harmonic signal, which was measured using a TE-cooled detector (Electro-Optical Systems Inc. S-050-TE2-H) fed into a lock-in amplifier (Stanford Research Systems, SR830).

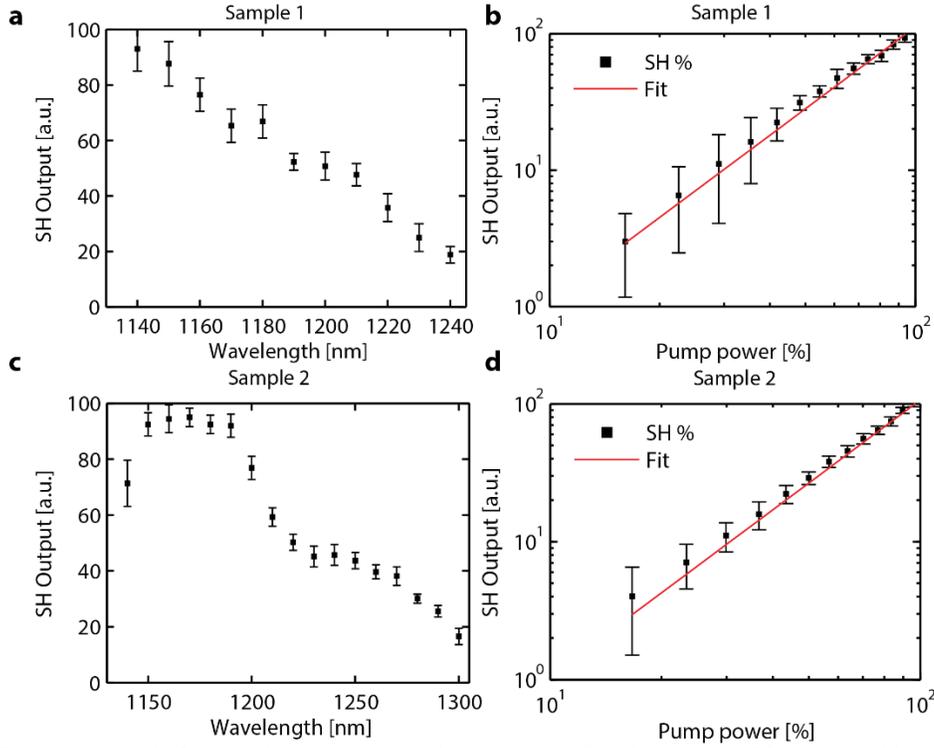

Fig. 2. Experimental results for transmitted second harmonic output. (a),(c) Wavelength scans of Samples 1 and 2, (b),(d) Power scans of Samples 1 and 2.

The transmitted SH signals measured from Samples 1 and 2 through wavelength and power scans are shown in Figure 2. Power scans, conducted near the fundamental of the electric resonance at 1170 nm, follow a second-order power law (shown in red in Figure 2b,d), ensuring detected signal was indeed at the second harmonic. Wavelength scans (Figure 2a,c) were conducted for constant 5 mW pump power. For Sample 1, we have observed a relative maximum at 1140 nm, which was where the SH electric resonance and fundamental magnetic resonances were matched. SH conversion efficiency ($\eta_{eff} = \hat{P}_{SH} / \hat{P}_{FF}$, where $\hat{P}_{SH}$ and $\hat{P}_{FF}$ are the SH and fundamental peak powers) is estimated to be $\sim 1.32 \times 10^{-10}$, or equivalently $1.2 \times 10^4$ photons/s are emitted by the metasurface. The SH efficiency trails off to noise level (~15 arb. un.) with no other features as we move off the aligned resonances. Due to limitations of the laser system, measurements for wavelengths below 1140 nm could not be taken. Below 1140 nm, the OPA in the laser setup generates components at 1240 nm, which would greatly skew the results for Sample 2.

For Sample 2, where the resonances were slightly detuned, we observe distinct peaks for both electric and magnetic resonances, however the SH signal observed from the electric resonance at the second harmonic is about two times stronger. SH conversion efficiency achieved at electric resonance is $\sim 1 \times 10^{-10}$, corresponding to emission of $9.08 \times 10^3$ photons/s. Hence, we clearly observe a 30% reduction of SH efficiency when the magnetic and electric resonances are detuned compared to matched resonances in Sample 1.

## 4. Simulations

This result is contradictory to most other measurements made using magnetic resonances, as they usually show much stronger enhancement to the nonlinear process. To elucidate this disparity, SH signal generated by our metasurface was simulated using finite-element frequency domain techniques using commercial software (COMSOL Multiphysics).

The second harmonic signal generated by the metasurface was simulated using a second-order polarization for the silver, which is dependent on silver's $\chi^{(2)}$ tensor components. Under Kleinmann symmetry, only three components of $\chi^{(2)}$ remain, specifically $\chi^{(2)}_{nnn} = 1.28 \times 10^{-20}$ m/V, $\chi^{(2)}_{ntt} = 1 \times 10^{-22}$ m/V, and $\chi^{(2)}_{ttn} = -8 \times 10^{-22}$ m/V, where $n$ and $t$ refer to components normal and tangential to the surface [39].

Second harmonic fields are solved for using the following equation [40]
$$\nabla \times \nabla \times E_2 - k_2^2 E_2 = \mu_0 \omega_2^2 P_2$$
where $E_2, k_2, \omega_2$ are the second-harmonic electric field, wave vector and frequency, and $P_2$ is the second-order polarization, which has the form [40]
$$P_2 = \epsilon_0 \chi_{ttn}^{(2)} E_{1,x} E_{1,z} \hat{x} + \epsilon_0 \chi_{ttn}^{(2)} E_{1,x} E_{1,z} \hat{y} + \epsilon_0 \left( \chi_{ntt}^{(2)} E_{1,x}^2 + \chi_{nnn}^{(2)} E_{1,z}^2 \right) \hat{z}$$
where $E_{1,x}$ and $E_{1,z}$ are the x- and z- components of the fundamental electric field. Simulated SH output is first normalized for Sample 1 according to SH efficiency at matched resonances and at noise level. The same noise level normalization is used for Sample 2 as well. Simulated SH output for both samples are shown in Figure 3a and Figure 3b (solid red line). Figure 3a
shows a maximum SH output for Sample 1 at 1140 nm as expected, which is where fundamental of electric resonance and magnetic resonance are matched. Figure 3b shows evident SH signal due to electric resonance at 1140 nm, but shows magnetic resonance to provide twice as much enhancement at 1240 nm. This result is in accordance with many of the previous experimental works, which have shown the magnetic resonance to provide much stronger enhancement of SHG. Confirming our experimental results, in simulations, we observe a 39% enhancement in SHG due to aligned resonances in Sample 1, compared to electric resonance in Sample 2.

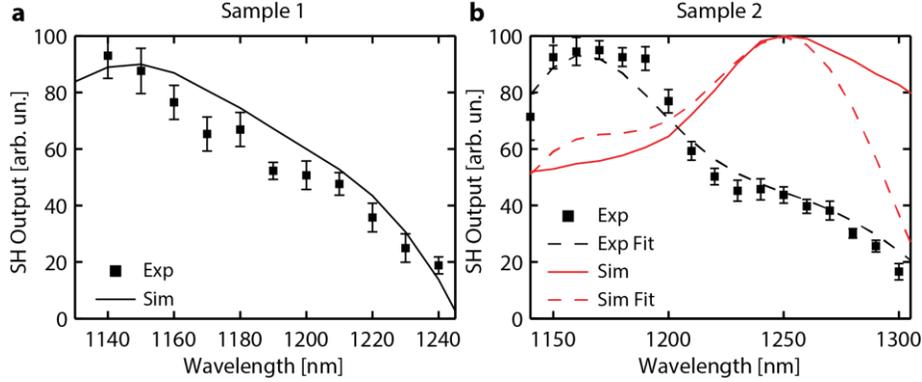

Fig. 3. (a) Match of experiments (black squares) and finite-element simulations (black line) for Sample 1, (b) Finite-element simulations (shown in solid red) that do not account for roughness in the structure and show the magnetic resonance at 1250 nm to have stronger contribution to SHG than the electric resonance at 1140 nm. Due to the computational complexity of simulating roughness for 3D structures by finite-element methods, we use a simplified Gaussian model to account for roughness and confirm our experimental results. The Gaussian "roughness" model is first tested against the finite-element simulations (compare dashed red vs. solid red) and then applied to fit the experiments (compare dashed black vs solid squares).

The disparity between experimental and simulated strengths of the isolated magnetic resonance can be explained by studying the role of roughness of the structure. As mentioned earlier, the magnetic resonance in the metasurface is highly sensitive to the thickness of the alumina spacer layer, while the electric resonance is quite robust, as seen in Figure 1d. Hence even 1-2 nm roughness in the metasurface will lead to significant inhomogeneous broadening of the second harmonic signal generated by the magnetic resonance, while the electric resonance would not be affected. Inhomogeneous broadening in linear spectra due to structural imperfections has been discussed previously in many experimental works [41-43].

We confirm the existence of inhomogeneous broadening in our metasurface by applying a 2-Gaussian fit for SH efficiency for both experiments and simulations. The Gaussian function of the form $f(\lambda) = a e^{-\frac{(\lambda-\mu)^2}{2\sigma^2}}$ is used for the electric and magnetic resonances. Gaussian functions are normalized using the same procedure as previously described for the nonlinear simulations. An electric resonance Gaussian at 1150 nm with σ = 42 nm is maintained for both experiments and simulations, while a Gaussian for the magnetic resonance at 1250 nm is inhomogeneously broadened from σ = 47 nm to σ = 80 nm to fit finite-element simulations and experiments, i.e. the integral of the spectral line shape is maintained. Experimental and simulation curves are shown in Figure 3, accompanied by respective Gaussian fits, which show a strong match, lending credence to our model of inhomogeneous broadening for experimentally observed spectra.

Since the magnetic resonance has a sensitivity of ~20nm/nm of Alumina in the spacer layer, we can calculate the approximate level of roughness in our structure by looking at the broadening of the SH output at the magnetic resonance. Since the resonance broadens from 47 nm FWHM to 80 nm FWHM between simulations and experiments, we can deduce that there is ~1.5 nm roughness in our spacer layer, which is similar to values presented in our previous experimental work [35,37].

## 5. Conclusions

We have studied a metasurface consisting of coupled silver nanostrips, exhibiting both electric and magnetic resonances for TM polarized light. We have experimentally observed that the electric resonance provides two times stronger peak enhancement to SHG than the magnetic resonance. We have also observed a 30% overall enhancement of SHG when electric and magnetic resonances were aligned. From simulations, we clearly see that magnetic resonances due to their sharper spectral features provide much stronge r enhancement for SHG than their electric counterparts. However, sharper spectral features come at a cost – they are easily affected by structural defects and roughness, which are unique to each structure and its fabrication processes. These imperfections can cause inhomogeneous broadening in experiments, thereby reducing peak enhancement values for nonlinear processes. Efficient wavelength converters at nanometer length scales, such as this device, are required for various integrated optics applications as well as nonlinear optical microscopy. An added benefit of such nanometer-thick devices are relaxed phase-matching conditions, allowing for low-dispersion and broadband wavelength conversion.

**Acknowledgements**

Authors acknowledge support from the US Army Research Office (ARO) Multidisciplinary University Research Initiative (MURI), grant #56154-PH-MUR (W911NF-09-1-0539), ARO grant 63133-PH (W911NF-13-1-0226), and the Air Force Office of Scientific Research (AFOSR) MURI grant FA9550-14-1-0389.

**References and links**

1. J. B. Pendry, D. Schurig, and D. R. Smith, "Controlling electromagnetic fields, " Science **312**, 1780-1782 (2006)
2. D. R. Smith, J. B. Pendry, and M. C. Wiltshire, "Metamaterials and negative refractive index," Science **305**, 788-792 (2004)
3. T. Ergin, N. Stenger, P. Brenner, J. B. Pendry, and M. Wegener, "Three-dimensional invisibility cloak at optical wavelengths," Science **328**, 337-339 (2010)
4. N. Yu, P. Genevet, M. A. Kats, F. Aieta, J.-P. Tetienne, F. Capasso, Z. Gaburro, "Light propagation with phase discontinuities: generalized laws of reflection and refraction," Science **334**, 333-337 (2011)
5. D. Gramotnev and S. I. Bozhevolnyi, "Plasmonics beyond the diffraction limit," Nature Photonics **4**, 83-91 (2010)
6. M. Kauranen and A. V. Zayats, "Nonlinear plasmonics," Nature Photonics **6**, 737-748 (2012)
7. P. Franken, A. Hill, C. Peters and G. Weinreich, "Generation of optical harmonics," Phys. Rev. Lett. **7**, 118-119 (1961)
8. N. Zheludev and V. Emel'yanov, "Phase matched second harmonic generation from nanostructured metallic surfaces," Journal of Optics A: Pure and Applied Optics **6**, 26 (2004)
9. W. Cai and V. M. Shalaev, "Optical Metamaterials," Vol. 10 (Springer, 2010)
10. C. Ciracì, E. Poutrina, M. Scalora, and D. R. Smith, "Second-harmonic generation in metallic nanoparticles: clarification of the role of the surface," Phys. Rev. B. **86**, 115451 (2012)
11. K. Thyagarajan, S. Rivier, A. Lovera, and O. J. F. Martin, "Enhanced second-harmonic generation from double resonant plasmonic antennae," Opt. Exp. **20**, 12860-12865 (2012)
12. G. F. Walsh and L. Dal Negro, "Enhanced second harmonic generation by photonic-plasmonic Fano-type coupling in nanoplasmonic arrays," Nano Letters **13**, 3111-3117 (2013)
13. J. Berthalot, G. Bachelier, M. Song, P. Rai, G. Colas des Francs, A. Dereux, and A. Bouhelier, "Silencing and enhancement of second-harmonic generation in optical gap antennas," Opt. Exp. **20**, 10498-10508 (2012)
14. B. K. Canfield, H. Husu, J. Laukkanen, B. Bai, M. Kuittinen, J. Turunen, and M. Kauranen, "Local field asymmetry drives second-harmonic generation in noncentrosymmetric nanodimers," Nano Letters **7**, 1251-1255 (2007)
15. A. Slablab, L. Le Xuan, M. Zielinski, Y. de Wilde, V. Jacques, D. Chauvat, and J.-F. Roch, "Second-harmonic generation from coupled plasmon modes in a single dimer of gold nanosphers," Opt. Exp. **20**, 220-227 (2012)
16. S. Palomba, M. Danckwerts, and L. Novotny, "Nonlinear plasmonics with gold nanoparticle antennas," Journal of Optics A: Pure and Applied Optics **11**, 114030 (2009)
17. H. Aouani, M. Navarro-Cia, M. Rahmani, T. P. H. Sidiropoulos, M. Hong, R. F. Oulton, and S. A. Maier, "Multiresonant broadband optical antennas as efficient tunable nanosources of second harmonic light," Nano Letters **12**, 4997-5002 (2012)
18. K. Thyagarajan, J. Butet, and O. J. F. Martin, "Augmenting second harmonic generation using fano resonances in plasmonic systems," Nano Letters **13**, 1847-1851 (2013)
19. J. Butet, I. Russier-Antoine, C. Jonin, N. Lascoux, E. Benichou, and P.-F. Brevet, "Effect of the dielectric core and embedding medium on the second harmonic generation from plasmonic nanoshells: tunability and sensing," The Journal of Physics Chemistry C **117**, 1172-1177 (2013)
20. Y. Pu, R. Grange, C.-L. Hsieh, and D. Psaltis, "Nonlinear optical properties of core-shell nanocavities for enhanced second-harmonic generation," Phys. Rev. Lett. **104**, 207402 (2010)

21. M. A. Vincenti, S. Campione, D. de Ceglia, F. Capolino, and M. Scalora, "Gain-assisted harmonic generation in near-zero permittivity metamaterials made of plasmonic nanoshells, "New Journal of Physics **14**, 103016 (2012)
22. M. W. Klein, C. Enkrich, M. Wegener, and S. Linden, "Second-harmonic generation from magnetic metamaterials," Science **313** 502-504 (2006)
23. M. W. Klein, M. Wegener, N. Feth, and S. Linden, "Experiments on second- and third- harmonic generation from magnetic metamaterials," Opt. Exp. **15**, 5238-5247 (2007)
24. S. Linden, F. B. P. Niesler, J. Förstner, Y. Grynko, T. Meier, and M. Wegener, "Collective effects in second-harmonic generation from split-ring resonator arrays," Phys. Rev. Lett. **109**, 015502 (2012)
25. N. Feth, S. Linden, M. W. Klein, M. Decker, F. B. P. Niesler, Y. Zeng, W. Hoyer, J. Liu, S. W. Koch, and J. V. Moloney, "Second-harmonic generation from complementary split-ring resonators," Opt. Lett. **33**, 1975-1977 (2008)
26. C. Ciracì, E. Poutrina, M. Scalora, and D. R. Smith, "Origin of second-harmonic generation enhancement in optical split-ring resonators," Phys. Rev. B **85**, 201403 (2012)
27. N. Segal, S. Keren-Zur, N. Hendler, and T. Ellenbogen, "Controlling light with metamaterial-based nonlinear photonic crystals," Nature Photonics **9**, 180-184 (2015)
28. K. O'Brien, H. Suchowski, J. Rho, A. Salandrino, B. Kante, X. Yin, and X. Zhang, "Predicting nonlinear properties of metamaterials from the linear response," Nature Materials **14**, 379-383 (2015)
29. Y. Zhang, N. K. Grady, C. Ayala-Orozco, and N. J. Halas, "Three-dimensional nanostructures as highly efficient generators of second harmonic light," Nano Letters **11**, 5519-5523 (2011)
30. I. Kolmychek, E. A. Mamonov, A. Y. Bykov, T. V. Murzina, S. Kruk, D. N. Neshev, M. Weismann, N. C. Panoiu, and Y. S. Kivshar, "Nonlinear-Optical Studies of Magnetic Dipole Metamaterials," in *Frontiers in Optics 2015*. (Optical Society of America) FTh1D.2asdf
31. S. S. Kruk, M. Weismann, A. Bykov, E. Mamonov, I. Kolmychek, T. Murzina, N. Panoiu, D. N. Neshev, and Y. S. Kivshar, in *CLEO:QELS Fundamental Science*, (2015), FM1C–4.
32. J. Lee, M. Tymchenko, C. Argyropoulos, P.-Y. Chen, F. Lu, F. Demmerle, G. Boehm, M.-C. Amann, A. Alu, and M. A. Belkin, "Giant nonlinear response from plasmonic metasurfaces coupled to intersubband transitions," Nature **511**, 65-69 (2014)
33. B. Metzger, L. Gui, J. Fuchs, D. Floess, M. Hentschel, and Harald Giessen, "Strong Enhancement of Second Harmonic Generation by Plasmonic Resonances at the Second Harmonic Wavelength," Nano Letters **15**, 3917-3922 (2015)
34. J. Butet, K. Thyagarajan, and O. J. F. Martin, "Ultrasensitive optical shape characterization of gold nanoantennas using second harmonic generation," Nano Letters **13**, 1787-1792 (2013)
35. W. Cai, U. K. Chettiar, A. V. Kildishev, and V. M. Shalaev, "Metamagnetics with rainbow colors," Opt. Exp. **15**, 3333-3341 (2007)
36. A. V. Kildishev, "Modeling nonlinear effects in 2D optical metamagnetics," Metamaterials **4**, 77-82 (2010)
37. V. P. Drachev, U. K. Chettiar, A. V. Kildishev, H.-K. Yuan, W. Cai, and V. M. Shalaev, "The Ag dielectric function in plasmonic metamaterials," Opt. Exp. **16**, 1186-1195 (2008)
38. X. Ni, Z. Liu, and A. V. Kildishev, "PhotonicsDB: Optical Constants," *Nanohub.org* (2007)
39. K. O'Donnell and R. Torre, "Characterization of the second-harmonic response of a silver-air interface," New Journal of Physics **7**, 154 (2005)
40. R. W. Boyd, *Nonlinear Optics* (Academic Press, 2003)
41. N. Papasimakis, V. A. Fedotov, Y. H. Fu, D. P. Tsai and N. I. Zheludev, "Coherent and incoherent metamaterials and order-disorder transitions," Phys. Rev. B **80**, 041102 (2009)
42. C. Rockstuhl, T. Zentgraf, H. Guo, N. Liu, C. Etrich, I. Loa, K. I. Syassen, J. Kuhl, F. Lederer, and H. Giessen, "Resonances of split-ring resonator metamaterials in the near infrared," Appl. Phys. B **84**, 219-227 (2006)
43. S. Chakrabarti, S. A. Ramakrishna, and H. Wanare, "Coherently controlling metamaterials," Opt. Exp. **16**, 19504-19511 (2008)

## Appendix: Supplementary Information

### 1. Field Maps and Transmission under 45° Illumination

Figures 1 and 2 below show the field maps and transmission of Sample 1 and Sample 2 at the electric and magnetic resonances under 45° illumination. As can be seen in Figures 1c and 2c, the magnetic resonances do not shift from the resonance wavelengths under normal incidence (1140 nm and 1240 nm), as shown in the paper. While the spectrum for 45° illumination at the electric resonance wavelength (570 nm) is different from that of normal illumination, the field maps (Figures 1a and 2a) show electric dipole behavior at 570 nm with significant enhancement.

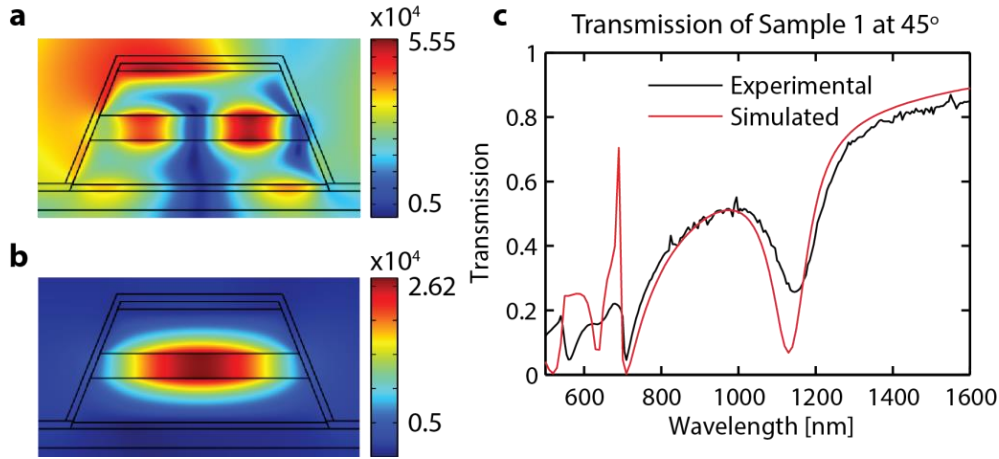

**Fig. 1.** Field maps of Sample 1 at (a) 570 nm (electric resonance) and (b) 1140 nm (magnetic resonance) and (c) transmission of Sample 1 at 45° incidence

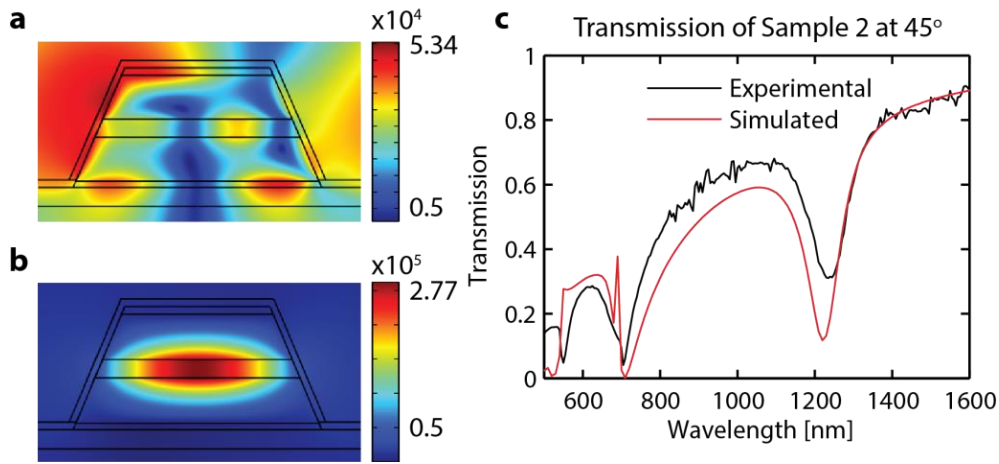

**Fig. 2.** Field maps of Sample 2 at (a) 570 nm (electric resonance) and (b) 1240 nm (magnetic resonance) and (c) transmission of Sample 2 at 45 degree° incidence

## 2. Laser Pulse Duration at Varying Wavelengths

Our Optical Parametric Amplifier (OPA) is fed by a Ti:Sapphire 800 nm femtosecond laser. The OPA outputs the selected IR wavelength, as well as residual 800 nm pump radiation. Using a Frequency-Resolved Optical Grating (FROG), we measured the pulse duration of this 800 nm pulse to be 98 fs ± 3 fs.
So, in order to ensure that all wavelengths from the OPA have similar pulse widths, we used a pump-probe setup (shown in Figure 3) using a BBO crystal to perform cross correlations between 800 nm pulses and our SHG pump pulses (1150, 1200 and 1250 nm) from the OPA. DM1 (Edmund Optics) is a dichroic mirror that transmits above 1050 nm and reflects between 790 and 1050 nm. DM2 (Thorlabs) is a dichroic mirror that reflects 750-820 nm. LF1 (Semrock, Inc.) is a 1064 long pass filter, and LF2 (Thorlabs) is a 700 long pass filter. A retroreflector setup was used to delay the 800 nm pulse. A CCD camera was used to visualize the spatial overlap of the two pulses on BBO crystal. The SFG signal was collected by a TE-cooled Si photodetector (ElectroOptical Systems) fed into a lock-in amplifier (Stanford Research Systems).

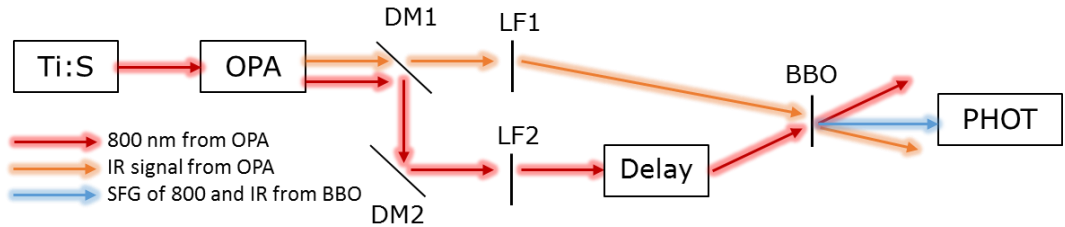

**Fig. 3.** Pump-probe setup used to measure pulse width of SHG pump beam. Red line refers to 800 nm light, orange light refers to IR SHG pump beam, and blue line refers to SFG signal from BBO crystal.

The cross-correlated sum frequency generation signals are depicted in Figure 4. The measured FWHM of these signals are 142 fs at 1150 nm, 138 fs at 1200 nm and 139 fs at 1250 nm. Assuming Gaussian pulse shape and using the standard relation $t_{SFG} = \sqrt{t_1^2 + t_2^2}$, we calculate the pulse widths of the OPA signals to be 102 fs for 1150 nm, 97 fs for 1200 nm, and 98 fs for 1250 nm. Hence, we confirm that pulse width does not vary significantly in wavelength range of interest (1130 nm – 1250 nm).

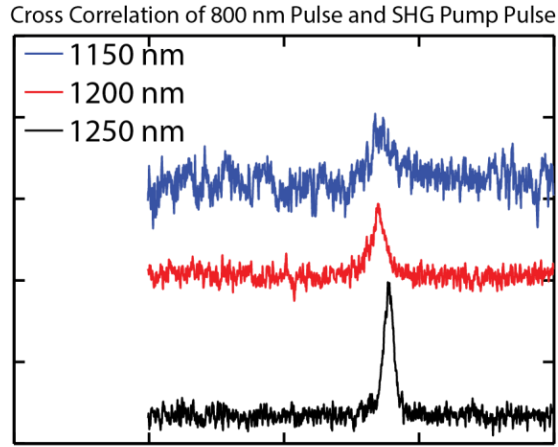

**Fig. 4.** Cross correlation signal of 800 nm and SHG pump pulses from Optical Parametric Amplifier (OPA).